\documentclass[12pt,onecolumn,draftclsnofoot]{IEEEtran}

\usepackage[encapsulated]{CJK}
\usepackage{ucs}
\usepackage[utf8x]{inputenc}
\usepackage[cmex10]{amsmath}
\usepackage{amsmath,amssymb,amscd,bbm,amsthm,mathrsfs,dsfont}
\usepackage{algorithmic,algorithm}
\usepackage{mdwmath}
\usepackage{mdwtab}
\usepackage{bm,upgreek}
\usepackage{cite}
\usepackage{rotating,graphics,psfrag,epsfig}
\usepackage{array}
\usepackage{booktabs}
\usepackage{indentfirst}
\usepackage{lipsum,fancyhdr,lastpage,refcount}
\usepackage{url}
\usepackage{fixltx2e,dblfloatfix}
\usepackage{blindtext}
\usepackage[T1]{fontenc}
\usepackage{color}
\usepackage[normalem]{ulem}
\usepackage{graphicx}
\usepackage{subcaption}

\IEEEoverridecommandlockouts

\let\oldnl\nl
\newcommand{\nonl}{\renewcommand{\nl}{\let\nl\oldnl}}

\begin{document}

\title{Computation Offloading in Beyond 5G Networks: A Distributed Learning Framework and Applications}

\author{\IEEEauthorblockN{Xianfu Chen, Celimuge Wu, Zhi Liu, Ning Zhang, and Yusheng Ji}

\thanks{X. Chen is with the VTT Technical Research Centre of Finland, Finland (e-mail: xianfu.chen@vtt.fi). C. Wu is with the Graduate School of Informatics and Engineering, University of Electro-Communications, Japan (e-mail: celimuge@uec.ac.jp). Z. Liu is with the Department of Mathematical and Systems Engineering, Shizuoka University, Japan (e-mail: liu@ieee.org). N. Zhang is with the Department of Electrical and Computer Engineering, University of Windsor, Canada (e-mail: ning.zhang@uwindsor.ca). Y. Ji is with the Information Systems Architecture Research Division, National Institute of Informatics, Tokyo, Japan (e-mail: kei@nii.ac.jp).}

\thanks{This work has been submitted to the IEEE for possible publication. Copyright may be transferred without notice, after which this version may no longer be accessible.}
}

\maketitle

\begin{abstract}

Facing the trend of merging wireless communications and multi-access edge computing (MEC), this article studies computation offloading in the beyond fifth-generation networks.
To address the technical challenges originating from the uncertainties and the sharing of limited resource in an MEC system, we formulate the computation offloading problem as a multi-agent Markov decision process, for which a distributed learning framework is proposed.
We present a case study on resource orchestration in computation offloading to showcase the potentials of an online distributed reinforcement learning algorithm developed under the proposed framework.
Experimental results demonstrate that our learning algorithm outperforms the benchmark resource orchestration algorithms.
Furthermore, we outline the research directions worth in-depth investigation to minimize the time cost, which is one of the main practical issues that prevent the implementation of the proposed distributed learning framework.

\end{abstract}


\section{Introduction}
\label{intr}

\begin{figure}[t]
  \centering
  \includegraphics[width=30pc]{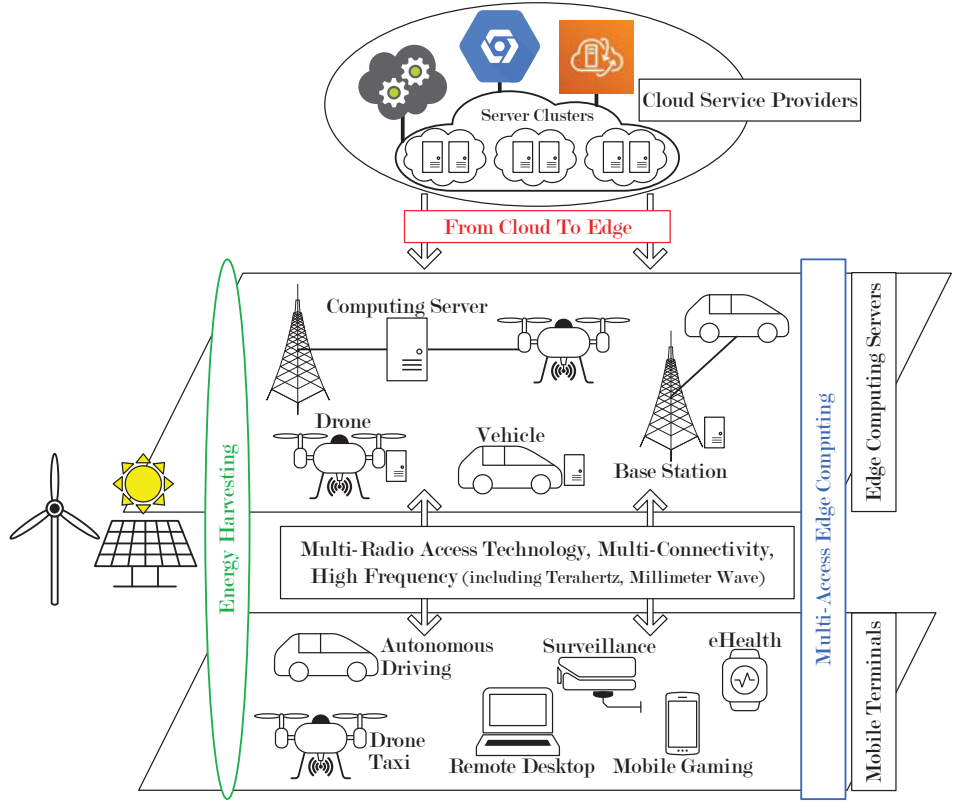}
  \caption{Computation offloading in the beyond 5G networks.}
  \label{MEC}
\end{figure}

Fifth-generation (5G) networks are ushering in the new era of wireless technology, which provides diverse services, including enhanced massive broadband, ultra-reliable and low-latency (URLL) communications and massive machine type communications \cite{ITUR17}.
Unfortunately, not all of these services can be achieved due to both the technical obstacles and the time pressure in deploying a new technology.
Beyond 5G networks are expected to enhance the 5G capabilities towards the support of seamless wireless connectivity with reliability and latency guarantees.
At the meantime, a multitude of mobile applications are emerging and gaining popularity, leading to a surge in computation demand.
However, the mobile terminals (MTs) are in general resource-constrained by the limited physical size.
To mitigate the burden from computation-intensive tasks with stringent URLL requirements, multi-access edge computing (MEC) is becoming one key technology by provisioning computing resources at the edge, in close proximity to MTs \cite{Abba18}.
The trend of merging wireless communications and MEC motivates the focus of computation offloading in beyond 5G networks in this article, as illustrated in Fig. \ref{MEC}.

The integration of MEC into beyond 5G networks holds the great potentials for improving the computation performance for MTs \cite{ETSI18}.
In such MEC systems, an MT decides whether the computation should be processed locally or offloaded to the edge computing servers for remote execution.
The design of computation offloading policies remains daunting.
Basically, the performance of computation offloading is predominantly bounded by two factors, namely, the wireless communication between an MT and the edge computing servers, and the remote execution at the edge computing servers, which involve a number of parameters \cite{Elba19}.
The wireless communication is performed by establishing the link with an edge computing server, controlling the input data size of the computation tasks to be offloaded, and selecting the frequency and the transmit power.
For the remote execution, computation resource allocation and task scheduling are of essence for computing efficiency.

To date, there have been extensive studies on computation offloading in MEC systems.
In \cite{You17}, You et al. formulated a convex optimization and a mixed-integer problems for the optimal resource allocation in a multiuser MEC system based on, respectively, time-division multiple access and orthogonal frequency-division multiple access.
A common drawback of the finite-time optimization is that the computation offloading parameters are considered to be irrelevant under different MEC system states, and hence the long-term performance cannot be sustained.
An infinite-time single-agent Markov decision process (MDP) was applied to investigate the dynamic computation offloading for an MEC system with wireless energy harvesting-enabled MTs in \cite{Mao16}, where the Lyapunov optimization technique constructs an approximately optimal solution.
To attain the optimal computation offloading policy, Xu et al. put forward an online reinforcement learning (RL) algorithm \cite{Xu17}.
However, the centralized decision-making limits the scalability of most existing RL-based algorithms.

When there are multiple decision-makers in an MEC system, the degree of cooperation plays a vital role in the design of computation offloading policies.
In the framework of a multi-agent MDP, an MT (i.e., a decision-maker) is regarded as an agent \cite{Rich98}.
At each time point of observation, the MEC system is in a state, and with the computation offloading policy, each agent takes an action, which is a decision-making in terms of a vector of the computation offloading parameters.
The MEC system responds by moving to a new state according to the probability distribution and sending feedback (i.e., the reward or cost signal) to each agent.
Due to the sharing nature of communication and computation resources, the actions from different agents are coupled.
With complete cooperation among the agents, the overhead incurred from inter-agent communications is overwhelming and prohibits the autonomous local actions at each agent.
Without any cooperation, the actions by each agent based on only the local information can be sub-optimal.
Therefore, the objective of this article is to develop a distributed learning framework for optimized computation offloading in beyond 5G networks, leveraging RL and supervised learning.

We organize the rest of this article as follows.
We highlight the technical challenges in computation offloading in beyond 5G networks.
To optimize the computation offloading performance, we propose a distributed learning framework.
We present a case study to show the potentials of an online distributed RL algorithm under the proposed learning framework.
We draw the conclusions and discuss the directions for future research.

\section{Challenges of Computation Offloading in Beyond 5G Networks}
\label{obst}

To facilitate efficient computation offloading in beyond 5G networks, the following inherent technical challenges need to be carefully addressed.

\subsection{Offloading Decision-Makings}

In addition to being locally processed by the MT in the MEC systems, a computation task can be fully or even partially offloaded to and executed by the edge computing servers.
Computation offloading is indeed a complex decision-making process, which accounts for the mobilities, the communication qualities, the computing capabilities and the resource availabilities.
The challenge arises from how to manage the computation offloading process.
Like the majority of studies, the offloading decision-making is formulated as a finite-time optimization problem given the communication as well as the computation resources.
The objective is either to minimize the energy consumption at each MT while keeping the latency acceptable for a specific computation task, or to find a tradeoff between the two.
However, the literature usually assumes a static MEC system environment.
In beyond 5G networks, the large-scale deployments make the environment much more complicated and it becomes more complex to re-formulate the offloading decision-making problem in accordance with the spatial and temporal dynamics.
Moreover, repeatedly collecting the global network information is costly.

\subsection{Resource Allocation}

After an offloading decision-making being made, the proper communication and computation resources have to be allocated accordingly, which is influenced by the edge computing sever selections, the task types and the latency requirements.
Specifically, if the task is separable, the latency requirement violation resulted from resource deficit can be avoided through parallel processing the computation at multiple edge computing servers. 
On the contrary, the latency requirement is restricted by the resource abundance of the selected edge computing server and the MT.
It is noteworthy that the edge computing server selection is highly dependent on the mobilities, because of which handovers occur.
To ensure the reliability requirement of task offloading, an MT initiates the handovers when it roams out of the coverage of the associated edge computing server or the experienced communication quality deteriorates.
In the beyond 5G networks, an enormous number of devices with heterogeneous communication and computing capabilities are deployed.
To deal with the heterogeneity and the numerous computation tasks with distinct reliability and latency requirements, the resource allocation asks for a flexible and distributed framework.

\section{Distributed Learning Framework}
\label{lfra}

In line with the preceding discussions, we develop a distributed learning framework for the computation offloading problem in beyond 5G networks, taking into consideration the relevance and the coupling among the decisions made by multiple MTs across the time.
Different from the assumption of a static environment, we adopt a multi-agent MDP to capture the dynamics that originates from the correlated uncertainties in an MEC system.
The uncertainties range from the variations in wireless communications between the MTs and the edge computing servers (e.g., the time-varying channel gains and the constantly changing network topologies) to the randomness in task computations (e.g., the sporadic task arrivals and the unpredictable energy sources) \cite{Sun19}.

\subsection{Overview of Multi-Agent MDPs}

\begin{figure}[t]
  \centering
  \includegraphics[width=26pc]{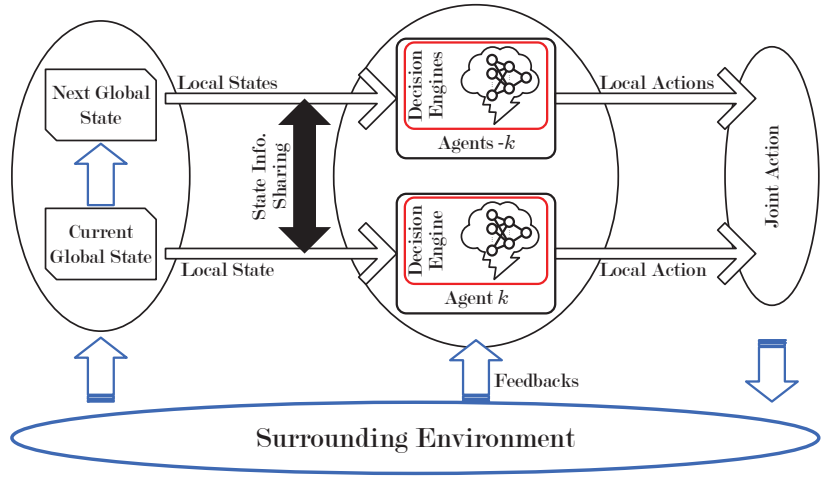}
  \caption{Diagram of a multi-agent Markov decision process, during which a set $\mathcal{K}$ of agents interact with the surrounding environment across the time.
  The decision engine of each agent $k \in \mathcal{K}$ outputs a local action with the input of not only the local state information, but also the state information from all the other agents $-k$.}
  \label{MMDP}
\end{figure}

As a straightforward extension of the single-agent MDP, a multi-agent MDP consists of multiple agents, jointly interacting with the surrounding environment, which is illustrated as in Fig. \ref{MMDP}.
At each time point, an agent is able to observe the local state that is only a local part of the global state.
Across the time, each agent shares the local state information with all the other agents.
The state information sharing enables the decision engine of an agent to take the global state information as the input and output a local action.
All the local actions by all the agents at a time point form a joint action, which leads to the transition of the surrounding environment from the current global state to the next global state, and meanwhile, a feedback is sent to each agent.
Notice that the global state transition follows the probability distribution over the next global states as a result of the joint action under the current global state.
The feedback associated with each agent is the reward or cost signal, which is used to evaluate the immediate effect of a joint action on the global state.
It is clear that the feedback indicates the inter-agent coupling among the local actions.
The goal of each agent is to solve a policy, which specifies the optimal local actions in different global states, so as to optimize the accumulated feedbacks in the long-run.

\subsection{Proposed Framework}

From the definition of a multi-agent MDP, we find that the obstacles to solving an optimal policy for an agent mainly lie in the state information sharing, the policy profile from other agents identifying the feedback and the knowledge of the global state transition probability.
RL techniques can be explored to learn the optimal policy without a priori knowledge of the global state transition probability.
However, on one hand, the feasibility of full state information sharing among the MTs is not obvious due to the large-scale deployment of a beyond 5G network.
On the other hand, the formulation of feedback received at each MT from the MEC system can be generally based on a combination of the consumed energy and the experienced latency, which is determined by the joint decision-makings from all MTs.
Under this context, the multi-agent MDP falls into a stochastic game.

\begin{figure}[t]
  \centering
  \includegraphics[width=28pc]{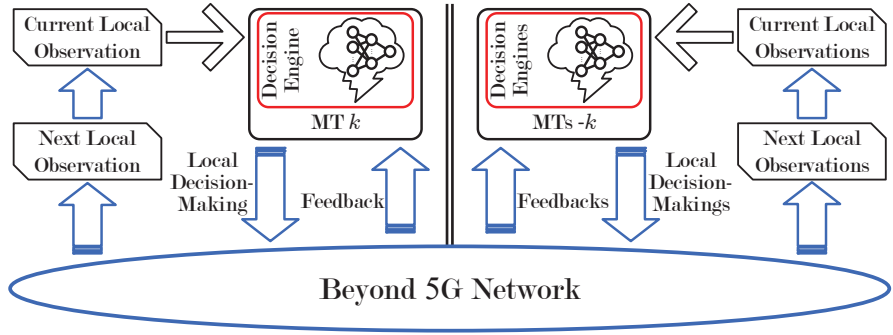}
  \caption{Proposed distributed learning framework.
  Each MT $k \in \mathcal{K}$ learns the optimal decision-making policy for computation offloading by making use of only the local observation information, where $\mathcal{K}$ denotes then the set of MTs in a beyond 5G network.}
  \label{DLF}
\end{figure}

As depicted in Fig. \ref{DLF}, this article proposes a distributed learning framework, whereby each MT learns the optimal policy and does not rely on the state information sharing with other MTs in the MEC system.
In the proposed framework, the local observation made by an MT at each current time point is composed of the current local state and the currently accessible contextual information.
In particular, the local state encapsulates the local awareness of the wireless communication variations and the task computation randomness.
The contextual information, which is the critical dimensionality of the local observation, allows an MT to behave in a distributed and autonomous manner.
At each time point, the local decision-making of each MT refers to the selection of a number of computation offloading parameters, while the emitted feedback from the MEC system is a measure of the energy consumption and the latency of task computation, as previously noted.
An MT is, therefore, enabled to independently learn an optimal computation offloading policy from the interactions with the MEC system, transforming the multi-agent MDP into a single-agent MDP.

This article introduces a novel categorization of contextual information as below.
\begin{enumerate}
  \item \emph{Conjecture} -- The offloaded computation tasks are executed by the edge computing servers.
      It is feasible for an MT that the cumulative distribution information of the historical decision-makings by other MTs in the MEC system can be proactively retrieved from the central network controller (e.g., a base station (BS) controller).
      Accordingly, each MT is capable of forming and updating conjectures on the behaviours of other MTs during the independent learning processes \cite{Chen13}.
  \item \emph{Abstraction} -- It is easy to see that a received feedback of an MT from the MEC system is related to a specific joint decision-making in a specific global state.
      That is, the classification of the feedback values into a finite number of intervals is equivalent to the abstraction of the local states of other MTs with bounded regrets \cite{Chen18}.
  \item \emph{Prediction} -- To exploit the side information in computation offloading from the past (e.g., the experienced latency and the energy consumption during wireless communications), a recurrent neural network architecture can be incorporated into the decision engine of each MT for an accurate local prediction of the global state at each time point \cite{Chen20}.
\end{enumerate}

Yet, in most practical MEC systems, each MT still faces a potential obstacle from the explosion of local observation and decision-making spaces.
The sufficient description of the uncertainties behind the local observations requires measuring various variables.
The recent advances in neural networks inspire us to empower the decision engine of an MT with a deep neural network, which has been proven to be a universal function approximator to represent the highly dimensional space of local observations \cite{Mnih15}.
For an MT, the number decision-makings grows exponentially as the number of computation offloading parameters increases.
Depending on the structure of the computation offloading problem, a linear decomposition approach can be employed to break the per-MT single-agent MDP into a series of simpler single-agent MDPs, each decision-making space of which is much smaller \cite{Chen18}.

\section{Resource Orchestration in Computation Offloading: A Case Study}
\label{appl}

In computation offloading, one interesting question is how to orchestrate the radio resources between the traditional communication and the computation services.
The resource orchestration is particularly challenging when taking into account the dynamics in the beyond 5G networks.
As a case study, we apply the proposed distributed learning framework for the resource orchestration in computation offloading.
We assume that a network operator (NO) implements a beyond 5G network, where the radio access network (RAN) is connected to a resource-rich edge computing server via fibre links.
The RAN sharing allows multiple wireless service providers (WSPs) to serve their respective MTs \cite{Gior20}.
Denote by $\mathcal{K}_j$ the set of MTs subscribed to an WSP $j$.
All MTs move across the discrete time in the service region covered by the RAN according to a Markov mobility model.
At each time point, the data packets and the number of computation tasks arriving to each MT follow a Poisson arrival process and a Markov chain, respectively.
The data packet arrivals get queued until transmissions, while the arrived tasks must be computed during a constant interval $\tau$ (in seconds) between two consecutive time points.

At each time point, the WSPs compete with each other for the limited frequency bands from the NO on behalf of the MTs.
If being granted a frequency band after the band competition, an MT proceeds to decide the number of data packets scheduled for transmissions and the number of computation tasks to be offloaded to the edge computing servers.
When the total number of data packet arrivals exceeds the queue size limit, overflows happen, resulting in packet drops.
The remaining computation tasks arrived at the MT are processed by the local central processing unit (CPU).
Given the MT mobilities, the time and energy consumptions during inter-BS handovers are fixed and cannot be optimized.
We further neglect the time consumption for sending back the computation outcomes from the edge computing server to the MT since the outcomes are typically much smaller than the input data size of a computation task.
Thus, we assume that the latency for task computation is kept to be $\tau$.
At each MT, the energy consumed by data transmissions and local CPU for task processing can then be calculated using the information of frequency bandwidth, channel gain to the associated BS, total size of scheduled packets as well as offloaded tasks, required CPU cycles to accomplish one input bit of the computation task and local CPU-cycle frequency.
The queueing delay and the overflows are the two important metrics to quantify the quality of traditional communication, for which we choose the queue length and the number of packet drops.

\subsection{Problem Formulation and Solution}

The target of each WSP $j$ is to design a control policy $\omega_j$, which optimizes the decision-makings $\mathbf{a}_j^t = \omega_j(\mathbf{s}^t)$ of frequency band competition, data packet scheduling and computation task offloading for the MTs under the global network states $\mathbf{s}^t$ across the discrete time $t = 1, 2, \cdots$, such that the long-term payoff is maximized.
The optimized long-term payoff of WSP $j$ is defined as $V_j(\mathbf{s}) = \textsf{E}[\sum_{t = 1}^\infty (\gamma)^{t  - 1} \cdot f_j^t | \mathbf{s}^1 = \mathbf{s}$.
Herein, $\mathbf{s}$ is the global network state at the current time point $t = 1$.
$\gamma \in [0, 1)$ is the discount factor.
$f_j^t = \sum_{k \in \mathcal{K}_j} \mu_k \cdot u_k^t - p_j^t$ is the immediate payoff of WSP $j$ at time point $t$, where $p_j^t$ is the payment to the NO for frequency band utilization, while $\mu_k$ and $u_k^t$ are, respectively, the positive weight and the immediate utility of an MT $k$.
The immediate utility $u_k^t$ of each MT $k$ is a reward signal of realized transmit energy consumption, local CPU energy consumption, queue length and packet drops after performing $(\mathbf{a}_j^t, \mathbf{a}_{- j}^t)$ under $\mathbf{s}^t$, where $\mathbf{a}_{- j}^t$ denotes the decision-makings from all the other WSPs.
The multi-agent MDP formulation of the resource orchestration problem is readily seen.
The coupling in decision-makings and the sharing of limited frequency bands among the WSPs makes the solving of an optimal control policy extremely hard.

\begin{figure}[t]
  \centering
  \includegraphics[width=29pc]{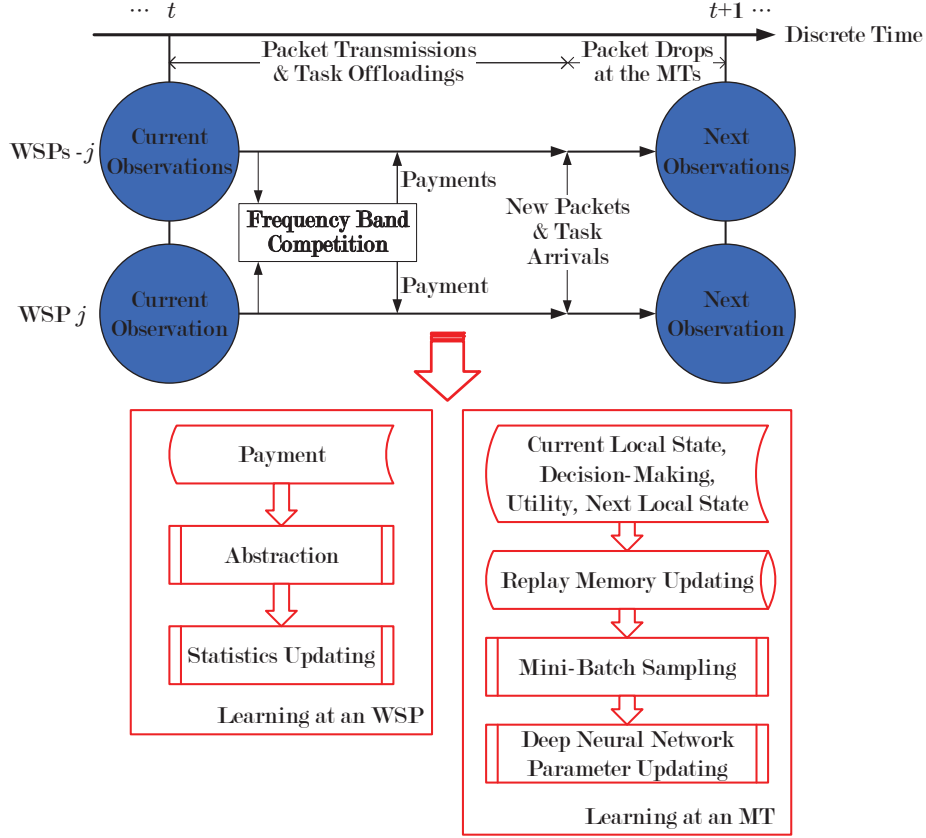}
  \caption{Flowchart of the online distributed deep RL algorithm.}
  \label{D_DRL}
\end{figure}

Based on our prior work as in \cite{Chen19}, we develop an online distributed multi-agent deep RL algorithm which decouples the decision-makings and enables the WSPs to independently learn the optimal control policy.
More specifically, we replace the local states of the MTs subscribed to other WSPs with the abstraction from the payment value classifications, which turns the original multi-agent MDP into a single-agent MDP for each WSP $j$ given by $V_j(\tilde{\mathbf{s}}_j) = \textsf{E}[\sum_{t = 1}^\infty (\gamma)^{t  - 1} \cdot f_j^t | \tilde{\mathbf{s}}_j^1 = \tilde{\mathbf{s}}_j]$, where the local observation $\tilde{\mathbf{s}}_j = (\mathbf{s}_j, c_j)$ of WSP $j$ at the current time point includes the local state information $\mathbf{s}_j$ of its MTs and the abstraction $c_j$.
For the purpose of reducing the dimensionality of a decision-making by WSP $j$, we linearly decompose $V_j(\tilde{\mathbf{s}}_j)$ through $V_j(\tilde{\mathbf{s}}_j) = \sum_{k \in \mathcal{K}_j} U_k(s_k) - U_j(c_j)$, where $U_k(s_k)$ and $s_k$ are the long-term utility and the current local state of each MT $k$, respectively, and $U_j(c_j)$ is the long-term payoff of WSP $j$.
The linear decomposition brings another advantage of letting the MTs to make local decisions of data packet scheduling and computation task offloading, while an WSP is responsible for frequency band competition only.
$U_j(c_j)$ can be easily learned by exploring the conventional RL algorithms.
To approach $U_k(s_k)$, we resort to a deep Q-learning algorithm to address the massive local state space and the unknown statistical network dynamics of MT $k$.
Fig. \ref{D_DRL} shows a flowchart of the developed online distributed deep RL algorithm for communication and computation resource orchestration employed by each WSP together with the subscribed MTs.

\subsection{Performance Evaluation}

We evaluate the developed distributed deep RL algorithm using numerical experiments based on TensorFlow.
The experiments simulate an RAN with 4 BSs distributed over a 2-kilometer by 2-kilometer square service region.
In the beyond 5G network, we assume 3 WSPs with each serving 6 MTs.
At each MT, we design a deep neural network with 2 hidden layers, and each layer contains 16 neurons.
A total of 11 frequency bands of equal bandwidth 500 KHz are shared among the WSPs.
The data sizes of a packet and a computation task are set to be 3000 bits and 5000 bits, respectively.
The numbers of task arrivals are random integers less than 6 across the time points, the interval $\tau$ of which is 0.01 seconds.
Running 1 bit of the task requires 737.5 CPU-cycles.
The CPU of each MT operates at the frequency of 2 GHz, while the maximum transmit power is 3 Watts.
In the payoff function of each WSP, the weight of each MT is chosen as 1.
For the sake of comparisons, we consider the following two benchmark algorithms, namely,
\begin{enumerate}
  \item \emph{Channel-Aware} -- The NO allocates the frequency bands to the MTs according to the channel gain information submitted by the WSPs. The priority of each MT is to transmit as many data packets as possible.
  \item \emph{Queue-Aware} -- The WSPs compete for the frequency bands in order to minimize the queue lengths and the packets drops of the MTs.
\end{enumerate}

\begin{figure*}
    \centering
    \begin{subfigure}[b]{18pc}
        \centering
        \includegraphics[width=16pc]{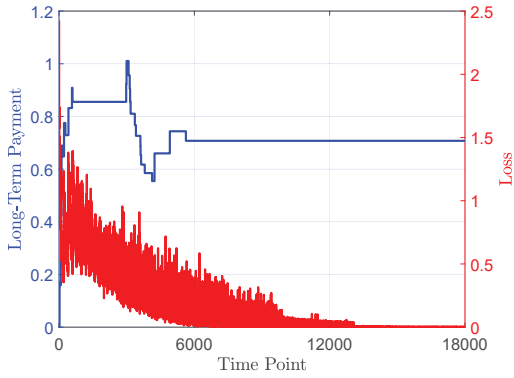}
        \caption{Convergence behavior of our developed online deep RL algorithm for resource orchestration.}
        \label{conv}
    \end{subfigure}
    \hfill
    \begin{subfigure}[b]{18pc}
        \centering
        \includegraphics[width=16pc]{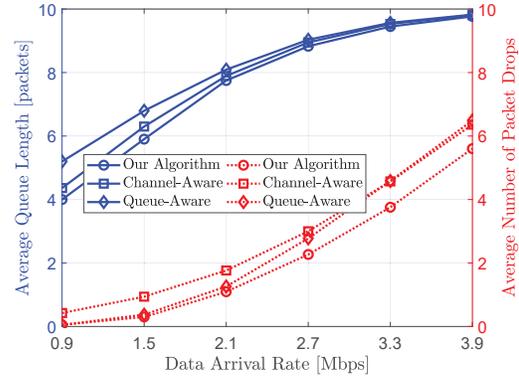}
        \caption{Average queue length and average number of packet drops across the discrete time.}
        \label{ener}
    \end{subfigure}
    \vskip\baselineskip
    \begin{subfigure}[b]{18pc}
        \centering
        \includegraphics[width=16pc]{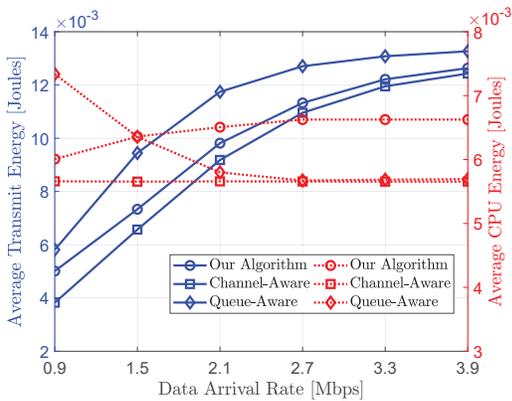}
        \caption{Average transmit energy and average local CPU energy consumptions across the discrete time.}
        \label{pack}
    \end{subfigure}
    \hfill
    \begin{subfigure}[b]{18pc}
        \centering
        \includegraphics[width=16pc]{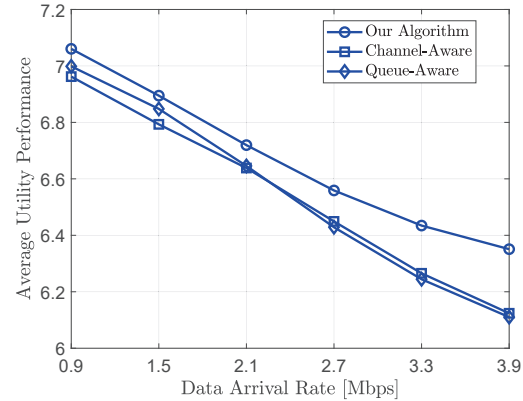}
        \caption{Achieved average utility performance for each MT across the discrete time.}
        \label{util}
    \end{subfigure}
    \caption{Experimental Results.}\label{resu}
\end{figure*}

The experimental results are showcased in Fig. \ref{resu}.
Fig. \ref{conv} examines the convergence in variations of the long-term payment for an WSP and the loss for an MT from updating, respectively, the abstraction statistics and the deep neural network parameters during the online learning process, where the data arrival rate of traditional communication for the MTs is 1.8 Mbps.
The curves tell that our developed online deep RL algorithm converges within around 13000 time points.
By changing the data arrival rates, we display the average performance for our algorithm and the two benchmark algorithms across the discrete time in Figs. \ref{ener}, \ref{pack} and \ref{util}.
We can observe that as the communication traffic load increases, both the average queue length and the average number of packet drops increase though the MTs consume more transmit energy on average for data packet transmissions.
Compared to the Channel-Aware algorithm, there exists a higher probability for an MT to be allocated a frequency band for transmitting data packets and offloading computation tasks when employing the Queue-Aware algorithm.
On the contrary, to reduce the packet drops by more data packet transmissions, our algorithm leaves more computation tasks to be locally processed, leading to an increase in average local CPU energy consumption.
Overall, our algorithm achieves better average utility performance than the Channel-Aware and the Queue-Aware algorithms.

\section{Conclusions and Future Directions}
\label{conc}

In this article, we concentrate on the investigation of computation offloading in the beyond 5G networks.
The dynamic uncertainties and the sharing of communication as well as computation resources necessitate the adoption of a multi-agent MDP to formulate the computation offloading problem when encountering multiple decision-makers.
To solve the multi-agent MDP problems, we propose a distributed learning framework, under which an MT is able to behave in a totally autonomous way with the awareness of the contextual information.
As a case study of the communication and computation resource orchestration, we demonstrate that our developed online deep RL algorithm under the distributed learning framework achieves a significant performance improvement in terms of average utility for each MT.

One important issue that prevents the practical implementations of the proposed distributed learning framework is the time cost in training the developed algorithms.
As can be seen from the case study, the convergence of an online learning algorithm is expensive.
In the following, we identify the research directions that deserve further investigation.
\begin{enumerate}
  \item \emph{Off-Policy Learning Algorithm Developments:} 
        The distributed RL algorithms require online interactions with the network elements, which is the main reason slowing down the training process.
        For an MT, the learning algorithm training aims at finding an optimal probability matching between a state and the actions.
        To minimize the training cost, one idea is to develop the off-policy learning algorithms that utilize the pre-collected online interaction data for offline training.
  \item \emph{Transfer Learning for Training Acceleration:} 
        The training of learning algorithms can be potentially enhanced if the historical learning experience can be leveraged.
        When the communication and computation requests from the network exhibit significantly temporal and spatial correlations, transferring the parameters of a converged learning algorithm among the MTs dramatically speeds up the training process, allowing faster decision-makings.
  \item \emph{Heterogeneity among MTs:}
        For the large-scale beyond 5G network deployments, there exist heterogeneities among MTs in the communication as well as computation capabilities and the contextual information availability.
        These heterogeneities have severe impacts on the stability of the training process if the MTs simultaneously learn the computation offloading policies.
        The development of distributed learning algorithms for such heterogeneous networks is demanding.
\end{enumerate}

\end{document}